\documentclass[pra,twocolumn,superscriptaddress,aps,amsmath,amssymb,nofootinbib]{revtex4}
\usepackage{graphicx}
\usepackage{dcolumn}
\usepackage{bm}
\usepackage{color}

\usepackage[section]{placeins}
\usepackage{graphicx,epsfig}
\usepackage{amsmath}
\usepackage{amsfonts}
\usepackage{braket}
\usepackage{fancyhdr}
\usepackage{hyperref}
\usepackage[utf8]{inputenc}
\usepackage[T1]{fontenc}
\usepackage{amsthm}
\usepackage{hhline}
\usepackage{multirow}
\usepackage[figuresleft]{rotating}
\def\beq{\begin{equation}}
\def\eeq{\end{equation}}
\def\beqa{\begin{eqnarray}}
\def\eeqa{\end{eqnarray}}

\begin{document}

\title{Spinodal instabilities of spin polarized asymmetric nuclear matter}

\date{\today}

\author{Artur Polls}
\address{Departament de F\'{i}sica Qu\`{a}ntica i Astrof\'{i}sica,
Universitat de Barcelona, Mart\'{i} i Franqu\`{e}s 1, 08028 Barcelona, Spain}
\address{Institut de Ci\`{e}ncies del Cosmos (ICCUB), Universitat de 
Barcelona, Mart\'{i} i Franqu\`{e}s 1, 08028 Barcelona, Spain}
\author{Isaac Vida\~na}
\address{Istituto Nazionale di Fisica Nucleare, Sezione di Catania, Dipartimento di Fisica ``Ettore Majorana'', Universit\`a di Catania, Via Santa Sofia 64, I-95123 Catania, Italy}

\begin{abstract}
We analyze the  spinodal instabilities of  spin polarized asymmetric nuclear matter at zero temperature for several configurations of the neutron and proton spins. The calculations are performed with the Brueckner--Hartree--Fock (BHF) approach using the Argonne V18 nucleon-nucleon potential plus a three-nucleon force of Urbana type. An analytical parametrization of the energy density, which reproduces with good accuracy the BHF results, is employed to determine the spinodal instability region. We find that, independently of the of the orientation of the neutron and proton spins, the  spinodal instability region shinks when the system is polarized, being its  size smaller smaller when neutron and proton spins are antiparallel than when they are oriented in a parallel way. We find also that the spinodal instability is always dominated by total density fluctuation independently of the degree of polarization of the system, and that restoration of the isospin symmetry in the liquid phase, {\it i.e.,} the so-called isospin distillation or fragmentation effect, becomes less efficient with the polarization of the system.
\end{abstract}
\maketitle
\section{Introduction}
\label{intro}

Phase transitions  are related to the thermodynamical instabilities that a physical system can present. 
Due to  the nature of the nucleon-nucleon interaction, which gives origin to an equation of state of the
Van der Waals type, a liquid-gas phase transition is expected to occur in nuclear matter \cite{bertsch83}. 
Multifragmentation in heavy-ion collisions, where highly excited composed nuclei are formed
in a gas of evaporated particles, can be used to study this transition. Results from these experiments
can be interpreted as the coexistence of a liquid and a gas phase  \cite{bondorf95,borderie04,gulminelli04,rivet05,trautmann05}.
Since nucleons can be either neutrons or protons, nuclear matter should be considered as a two-component fluid. Therefore,
it is expected that thermodynamical instabilities in nuclear matter give rise to a quite rich phase 
diagram \cite{barranco80,muller95,baran98,baran01,baran05}.  A lot of interest has been devoted  to define the nature of these instabilities. Usually, it has been argued that asymmetric nuclear matter presents two types 
of instabilities: a mechanical (or isoscalar)  instability associated with density fluctuations which conserve the proton fraction, and a chemical (or isovector) instability, 
related to fluctuations in the proton fraction, occurring at constant density. However, it was
demonstrated \cite{baran01,margueron03,chomaz04} that asymmetric nuclear matter presents in fact only one type of instability  
dominated by total density fluctuations which lead to a liquid-gas phase separation with restoration of the 
isospin symmetry in the liquid dense phase. This phenomena, where large droplets of  high density symmetric matter are formed in 
a background of a neutron gas with a small fraction of protons,  is known as  isospin distillation or fragmentation effect \cite{xu00}.

The stability conditions of isospin asymmetric nuclear matter against the liquid-gas phase transition have been systematically analyzed by using 
different approaches that include mean field calculations with effective forces of Skyrme or Gogny type \cite{margueron03,ducoin06,ducoin07a,ducoin07b}, 
relativistic mean field calculations using constant and density-dependent couplings parameters \cite{liu02,avancini04,providencia06,avancini06,santos08} or
Dirac--Brueckner--Hartree--Fock \cite{margueron07} and Brueckner--Hartree--Fock (BHF) \cite{vidana08} approaches with realistic nucleon-nucleon interactions.
In all these analysis it has been always considered that both protons and neutrons are spin saturated, {\it i.e.,} non polarized.  However, the presence of  strong magnetic fields, such as those estimated in neutron stars \cite{pacini67,gold68}, particularly in magnetars \cite{duncan92,thompson95,usov92,paczynski92},  or those predicted in noncentral heavy-ion collisions \cite{kharzeev06,kharzeev08,skokov09,mo13}, can induce the polarization of the neutron and proton spins. It is, therefore, interesting to extend the analysis of the stability conditions of nuclear matter to the spin polarized case.  To the best of our knowledge, this extended analysis has been only done using RMF models in Refs.\ \cite{rabhi09,fang17,avancini18}. In these works,  the effect of strong magnetic fields on the spinodal instabilities, the isospin distillation and the crust-core transition in neutron stars have been studied. The results of these studies show that sufficiently strong magnetic fields can significantly modify the extension of the unstable region. Multifragmentation experiments using polarized targets and projectile beams could potentially explore the modification of the unstable region and allow the study of a more complex nuclear matter phase diagram in which different phases with different spin and isospin content could coexist.

A general study of the stability conditions of spin polarized nuclear matter against a phase separation requires the analysis of the convexity of its free-energy density with respect to the partial densities of the four fluids that compose the system: neutrons and protons both with spin up and down. However, 
the degree of spin polarization of neutrons and protons in systems like neutron stars or in experiments with polarized targets and projectile beams is fixed by the magnetic field. Therefore, it is of interest to analyze the spinodal instabilities at different fixed values of neutron and proton spin polarizations, and explore their effect on the nature of the instabilities. In this work we perform this analysis at zero temperature using the BHF approach with the realistic Argonne V18 \cite{av18} nucleon-nucleon force supplemented with a three-nucleon force of the Urbana type \cite{uix1,uix2} which for the use in the BHF calculation is reduced to a two-body density dependent force by averaging over spatial, spin and isospin coordinates of the third nucleon.

The manuscript is organized in the following way. A brief review of the BHF approach for spin polarized asymmetric nuclear matter is made in Sec.\ \ref{sec:bhf}.  The stability criteria against phase separation for spin polarized matter are presented in Sec.\ \ref{sec2}.  Results are shown and discussed in Sec.\ \ref{sec3}. Finally, a summary and the main conclusions of this work are given in Sec.\ \ref{conclusions}.


\section{BHF approach of spin polarized asymmetric nuclear matter}
\label{sec:bhf}

Spin polarized asymmetric nuclear matter is an ideal infinite nuclear system composed of four different fermionic components: neutrons with spin up and down having densities $\rho_{n_\uparrow}$ and $\rho_{n_\downarrow}$, respectively, and protons with spin up and down with densities $\rho_{p_\uparrow}$ and $\rho_{p_\downarrow}$. The total density of the system is  
\begin{equation}
\rho=\rho_{n_\uparrow}+\rho_{n_\downarrow}+\rho_{p_\uparrow}+\rho_{p_\downarrow}\equiv\rho_n+\rho_p \ ,
\label{eq:totde}
\end{equation}
where $\rho_n$ ($\rho_p$) is the total density of neutrons (protons).
The isospin asymmetry of the system can be expressed by the asymmetry parameter $\beta=(\rho_n-\rho_p)/\rho$, while its degree of spin polarization can be characterized by the neutron and proton spin polarizations $S_n$ and $S_p$, defined as
\begin{equation}
S_n = \frac {\rho_{n_\uparrow} - \rho_{n_\downarrow}}{\rho_n}\ , \,\, S_p = \frac {\rho_{p_\uparrow}- \rho_{p_\downarrow}}{\rho_p} \, .
\label{eq:spinasy}
\end{equation}
Note that the values $S_n=S_p=0$ correspond to non-polarized matter ($\it i.e.,$ $\rho_{n\uparrow}=\rho_{n\downarrow}$ and $\rho_{p\uparrow}=\rho_{p\downarrow}$), whereas $S_n=\pm 1$ ($S_p=\pm 1$) means that neutrons (protons) are totally polarized, {\it i.e.,} all neutron (proton) spins are aligned along the same direction.

The single densities are related to the total one $\rho$ and the isospin and spin asymmetry parameters $\beta, S_n$ and $S_p$ through the equations
\begin{equation}
\rho_{n_\uparrow}=\frac{(1+S_n)(1+\beta)}{4}\rho\ , \,\, \rho_{n_\downarrow}=\frac{(1-S_n)(1+\beta)}{4}\rho\ ,
\label{eq:partden1}
\end{equation}
\begin{equation}
\rho_{p_\uparrow}=\frac{(1+S_p)(1-\beta)}{4}\rho\ , \,\, \rho_{p_\downarrow}=\frac{(1-S_p)(1-\beta)}{4}\rho\ .
\label{eq:partden2}
\end{equation}

Our many-body scheme starts with the construction of all the $G$ matrices that describe  the in-medium interaction of two nucleons  ($nn,np,pn$ and $pp$) for each one of the spin combinations ($\uparrow\uparrow,\uparrow\downarrow,\downarrow\uparrow$ and $\downarrow\downarrow$). The $G$ matrices are obtained by solving the well known Bethe--Goldstone equation
\begin{widetext}
\begin{eqnarray}
\langle \vec k_1 \tau_1\sigma_1; \vec k_2 \tau_2\sigma_2|G(\omega)|\vec k_3 \tau_3\sigma_3; \vec k_4 \tau_4\sigma_4\rangle
&=&
\langle \vec k_1 \tau_1\sigma_1; \vec k_2 \tau_2\sigma_2|V|\vec k_3 \tau_3\sigma_3; \vec k_4 \tau_4\sigma_4\rangle
+\sum_{ij}
\langle \vec k_1 \tau_1\sigma_1; \vec k_2 \tau_2\sigma_2|V|\vec k_i \tau_i\sigma_i; \vec k_j \tau_j\sigma_j\rangle \nonumber \\
&\times&
\frac{Q_{\tau_i\sigma_i,\tau_j\sigma_j}(\vec k_i,\vec k_j) }{\omega-E_{\tau_i\sigma_i}(\vec k_i)-E_{\tau_j\sigma_j}(\vec k_j)+i\eta}
\langle \vec k_i \tau_i\sigma_i; \vec k_j \tau_j\sigma_j|G(\omega)|\vec k_3 \tau_3\sigma_3; \vec k_4 \tau_4\sigma_4\rangle \ ,
\label{eq:gmat}
\end{eqnarray}
\end{widetext} 
where $\tau$ and $\sigma$ indicate, respectively, the  isospin ($n,p$) and spin  ($\uparrow,\downarrow$) projections of  the two nucleons in the initial, intermediate and final states, $\vec k$ are their respective linear momenta,
 $V$ is the bare nucleon-nucleon interaction (in our case the Argonne V18
plus the UIX three-body force reduced to a two-body density dependent one), $Q_{\tau_i\sigma_i,\tau_j\sigma_j}(\vec k_i,\vec k_j)$ is the Pauli operator which allows only intermediate states compatible with the Pauli principle, and $\omega$ is the sum of the non-relativistic energies of the interacting nucleons. 

The single-particle energy of a nucleon ($\tau=n,p$) with spin projection $\sigma=\uparrow, \downarrow$ and momentum $\vec k$ is given by $E_{\tau\sigma}(\vec k)=\frac{\hbar^2k^2}{2m_\tau}+U_{\tau\sigma}(\vec k)$, where the single-particle potential $U_{\tau\sigma}(\vec k)$ represents the mean field ``felt'' by the nucleon due to its interaction with the other nucleons of the system. In the BHF approach $U_{\tau\sigma}(\vec k)$ is calculated through the ``on-shell'' $G$ matrices 
\begin{widetext}
\begin{equation}
U_{\tau\sigma}(\vec k)=\sum_{\tau'\sigma'}\sum_{k'\leq k_{F}^{\tau\sigma}}
\langle \vec k \tau\sigma; \vec k' \tau'\sigma'|G(\omega=E_{\tau\sigma}(\vec k)+E_{\tau'\sigma'}(\vec k'))|\vec k \tau\sigma; \vec k' \tau'\sigma'\rangle_{\cal A} \ ,
\label{eq:spp}
\end{equation}
\end{widetext}
where a sum over the Fermi seas of neutron and protons with spin up and down is performed and the matrix elements are properly antisymmetrized when required. We note that the continuous prescription has been adopted when solving the Bethe--Goldstone equation. Once a self-consistent solution of Eqs.\ (\ref{eq:gmat}) and (\ref{eq:spp}) is obtained the total energy density can be easily obtained as
\begin{equation}
\varepsilon = 
\sum_{\tau\sigma}\int_0^{k\leq k_{F}^{\tau\sigma}}\frac{d^3k}{(2\pi)^3} \left(\frac{\hbar^2k^2}{2m_\tau}
+\frac{1}{2}U_{\tau\sigma}(\vec k) \right) \ .
\end{equation}
This quantity is obviously a function of the partial densities $\rho_{n_\uparrow}, \rho_{n_\downarrow},\rho_{p_\uparrow}$ and $\rho_{p_\downarrow}$ or, equivalently of the total density $\rho$, the isospin asymmetry $\beta$ and the spin polarizations $S_n$ and $S_p$. 

\begin{figure}[b]
\centering
\includegraphics[width=1.0 \columnwidth]{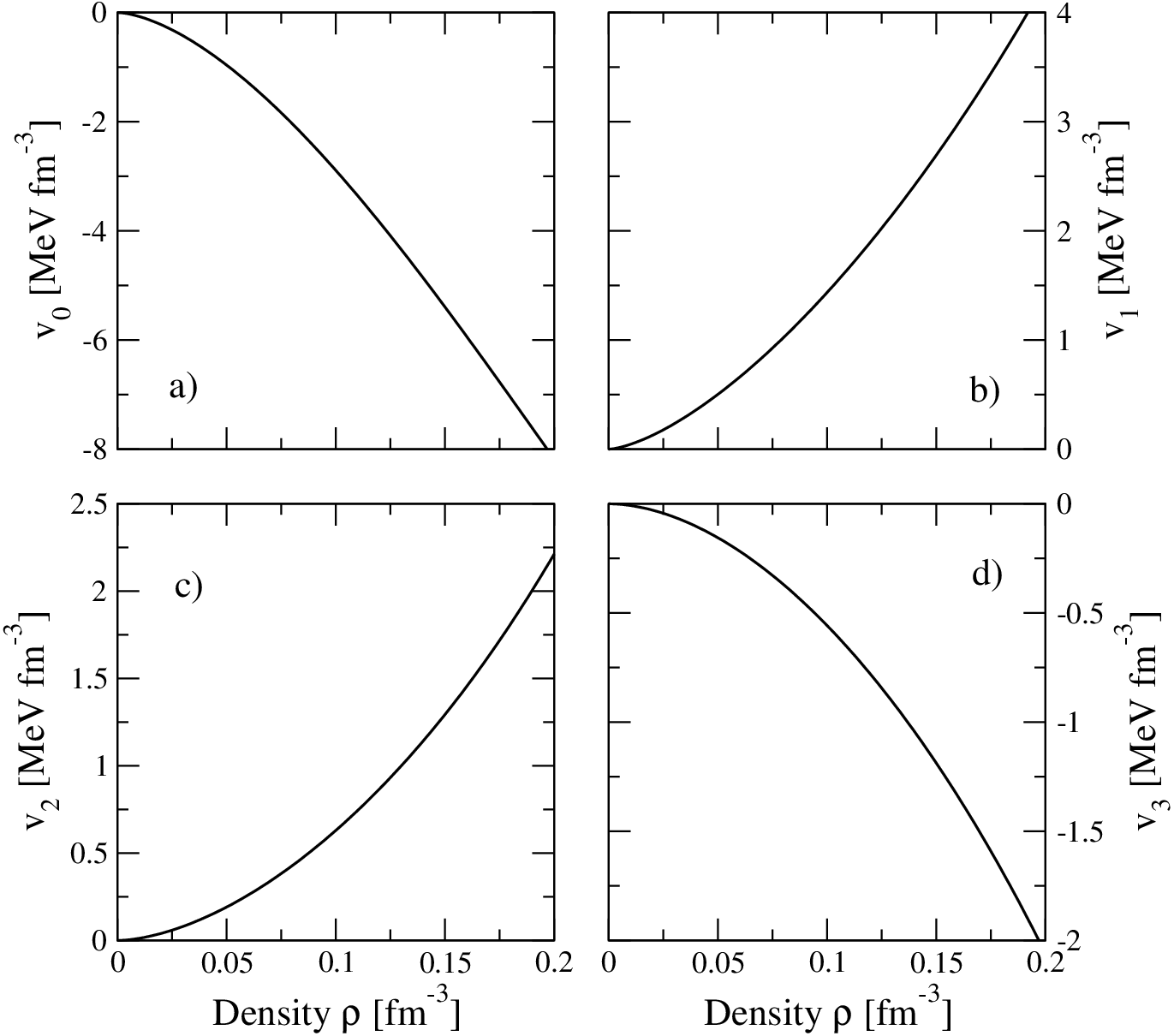}
        \caption{Density dependence of the coefficients $v_i$ $(i=0,\cdot\cdot\cdot,3)$.}
\label{fig:coeff}
\end{figure}

BHF calculations are in general quite expensive in terms of computational time. Therefore, from a practical point of view, it is very useful to have an analytical parametrization of the BHF energy density that allow us to determine the spinodal instability region in a fast way and, in addition, facilitates us the interpretation of the results.  In this work we use  the following energy density functional,  developed  by one of the authors in Ref.\ \cite{vidana02}, that parametrizes the BHF results for spin polarized asymmetric nuclear matter
\begin{eqnarray}
\varepsilon(\rho,\beta,S_n,S_p)&=&t(\rho,\beta,S_n,S_p)+v_0(\rho)+v_1(\rho)\beta^2 \nonumber \\
&+&v_2(\rho)(1+\beta)^2S_n^2+v_2(\rho)(1-\beta)^2S_p^2 \nonumber \\
&+&v_3(\rho)(1-\beta^2)S_nS_p \ ,
\label{eq:para}
\end{eqnarray}
where
\begin{eqnarray}
t(\rho,\beta,S_n.S_p)&=&\frac{3}{5}\frac{\hbar^2k_F^2}{2m}\frac{\rho}{4}\left[ (1+\beta)^{5/3} (1+S_n)^{5/3} \right. \nonumber \\
&+&(1+\beta)^{5/3} (1-S_n)^{5/3} \nonumber \\
&+&(1-\beta)^{5/3} (1+S_p)^{5/3} \nonumber \\
&+&\left.(1-\beta)^{5/3} (1-S_p)^{5/3}\right] 
\label{eq:kin}
\end{eqnarray}
is the kinetic energy density with $k_F=(3\pi^2\rho/2)$, and the coefficients $v_i(\rho)$ $(i=0,\cdot\cdot\cdot,3)$ have been determined by imposing the parametrization of Eq.\  (\ref{eq:para}) to reproduce the BHF results corresponding to the following four sets of values of $\beta$, $S_n$ and $S_p$: $(\beta=0, S_n=0,S_p=0)$, $(\beta=1, S_n=0,S_p=0)$, $(\beta=0, S_n=1, S_p=0)$ and $(\beta=0,S_n=1,S_p=1)$. The density dependence of the coefficients, assumed to be of the form
\begin{equation}
v_i(\rho)=a\rho^{\gamma}+b\rho^{\delta} \ , \,\,\,\,\, i=0, \cdot\cdot\cdot, 3 
\end{equation}
is shown in Fig.\ \ref{fig:coeff}. The set of parameters $a, \gamma, b$ and $\delta$  is given in Tab.\ \ref{t:tab1}.

\begin{table}[t]
\begin{center}
\begin{tabular}{c|cccc}
\hline
\hline
coefficient & $a$ & $\gamma$ & $b$ & $\delta$ \\
\hline
$v_0$        & $-118.92$ & $1.60$ & $484.31$ & $3.99$ \\
$v_1$        & $44.48$    & $1.49$  & $115.53$ & $3.73$ \\
$v_2$        & $30.86$   & $1.69$   & $159.35$  &  $ 4.13$ \\ 
$v_3$        & $-38.16$    & $1.83$   & $-265.51$ & $5.07$ \\
\hline
\hline
\end{tabular}
\end{center}
\caption{Set of parameters $a, \gamma, b$ and $\delta$ characterizing the density dependence of the coefficients $v_i$. The parameters $\gamma$ and $\delta$ are dimensionless whereas the units of $a$ and $b$ are MeV$\times$ fm$^{3\gamma-3}$ and MeV$\times$ fm$^{3\delta-3}$, respectively.}

\label{t:tab1}
\end{table}

\begin{figure*}[t]
\centering
\includegraphics[width=1.2 \columnwidth]{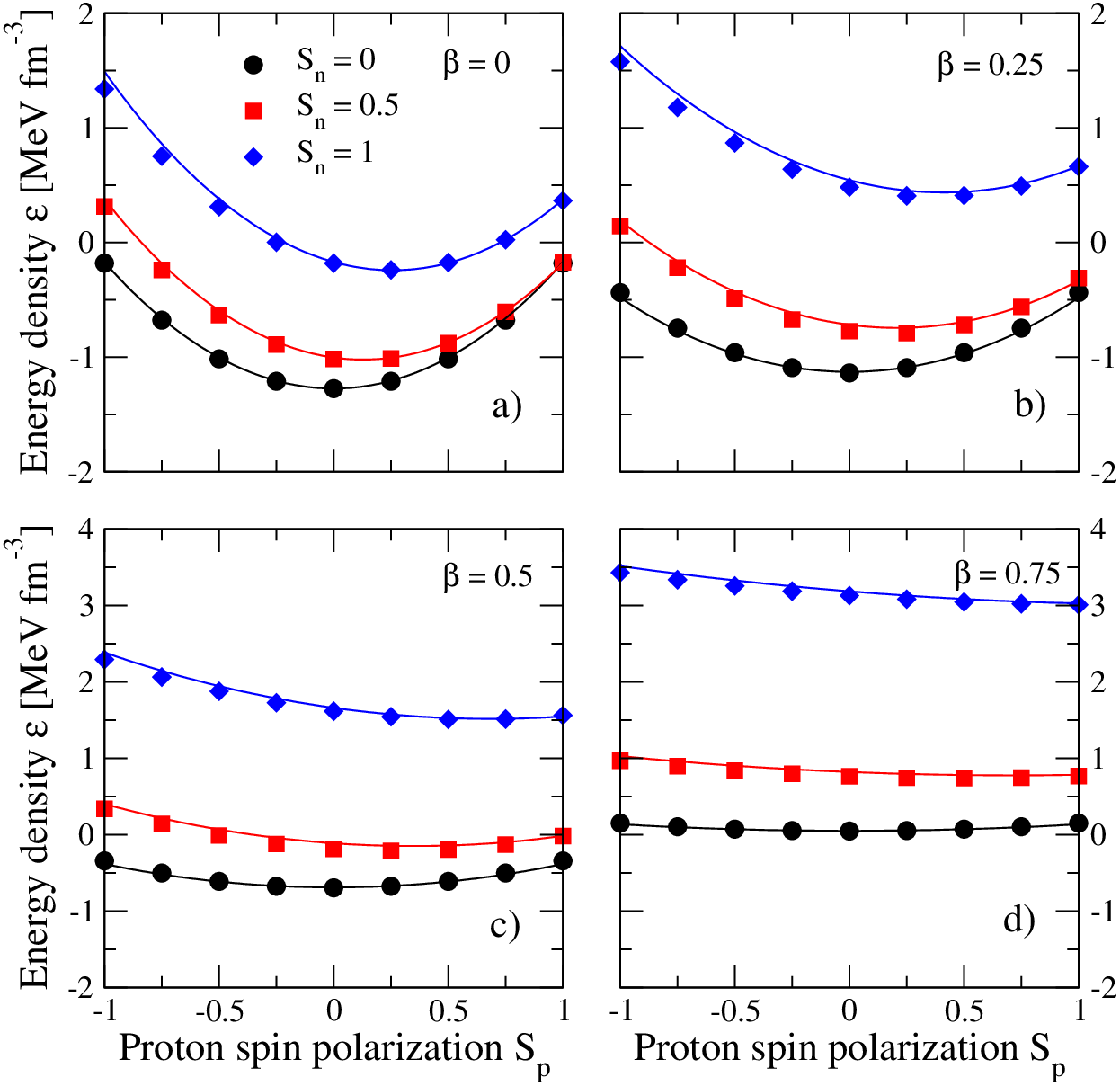}
        \caption{(Color online) Energy density at $\rho=0.1$ fm$^{-3}$ as a function of $S_p$ for different values of $\beta$ and $S_n$. Symbols show the result of the BHF calculation whereas solid correspond to those obtained from the parametrization of Eq.\ (\ref{eq:para}).}
\label{f:bhf_fit}
\end{figure*}

We note that the determination of these coefficients is not unique and we could have impose the parametrization to reproduce the BHF results for a different set of values of $\beta, S_n$ and $S_p$.  However, by choosing them in this way,  we get a parametrization that  reproduces with a good quality the results of the BHF calculations in a wide range of values of the isospin and spin asymmetry parameters, as it can be seen in Fig.\ \ref{f:bhf_fit}. Symbols show the results obtained from the BHF calculation whereas those obtained from the parametrization are reported by solid lines. As it can be seem from the figure the spin polarization and isospin asymmetry predicted by the microscopic calculation is well reproduced by the parametrization. The quality of the parametrization is quite good with deviations from the microscopic calculation of just a few percent only for values of $\beta, S_n$ and $S_p$ corresponding to the most isospin and spin asymmetric cases. It is interesting to observed that for fixed values of $\beta$ and $S_n$ (with $S_n\neq 0$), the minimum of the energy density occurs for a value of $S_p\neq 0$. However, we should note that this is not an indication of a ferromagnetic instability signaling a phase transition from the non-polarized state to a polarized one of lower energy, because the real ground state of the system is always the non-polarized one ($S_n=0, S_p=0$).

\begin{figure}[t]
\centering
\includegraphics[width=1.0 \columnwidth]{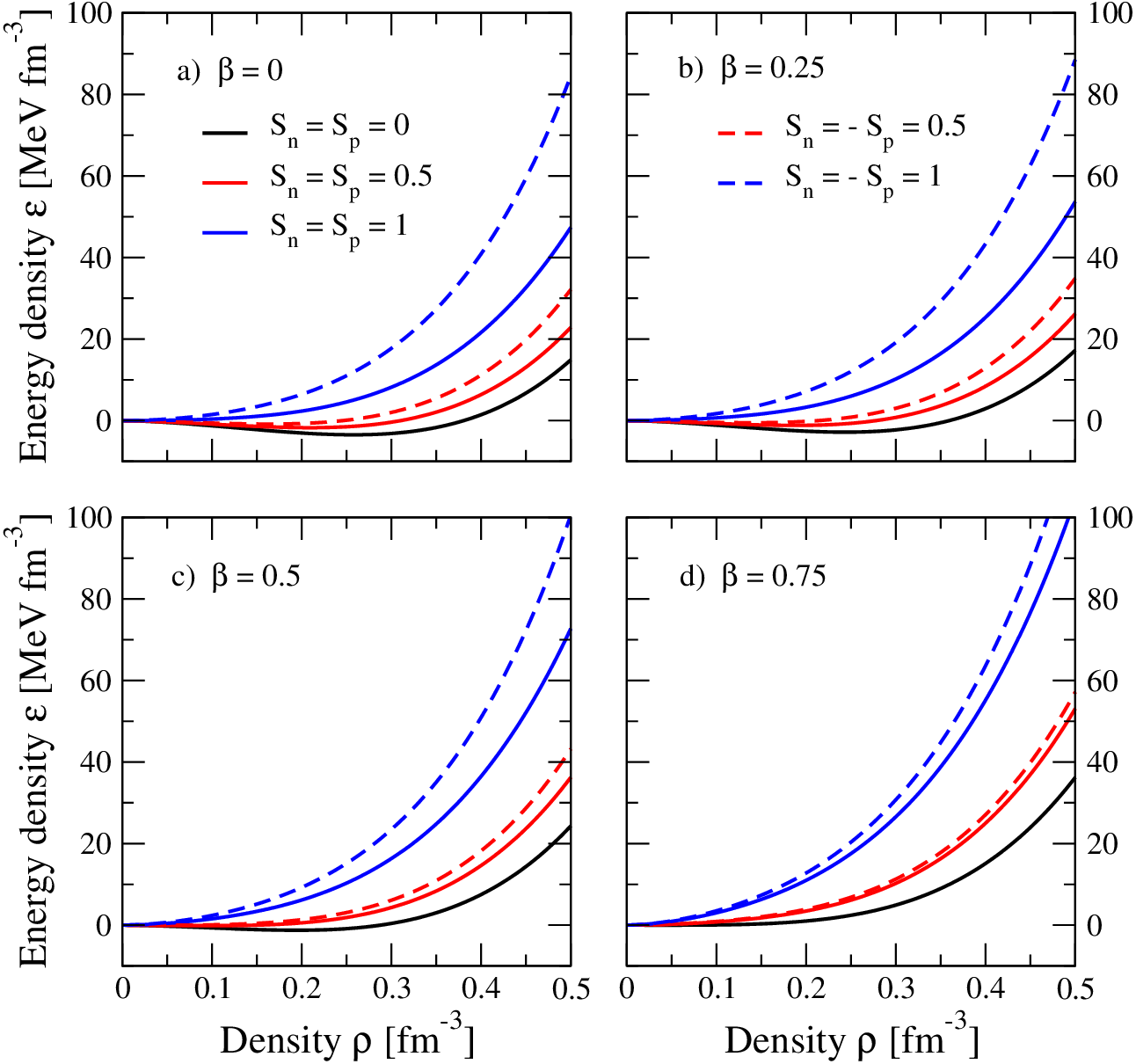}
        \caption{(Color online) Energy density as a function of the total density of the system for different values of $\beta$, $S_n$ and $S_p$. }
\label{f:edensplot}
\end{figure}

To finish this section, we show in Fig.\ \ref{f:edensplot} the energy density $\varepsilon$ as a function of the total density of the system for different values of 
$\beta$ and $S_n$ and $S_p$. We note that, the energy density of spin-polarized matter is always larger than that of non-polarized matter in the whole range of densities for any value of the isospin asymmetry. Furthermore, it increases when increasing the spin polarization of the system. Note also that this increase is even larger when the neutron and proton spins have an antiparallel orientation the system. A comprehensive explanation of the behavior of the energy density with the spin polarization was already given in Ref.\ \cite{vidana02}, and we refer the interested reader to this work for details.

\section{Stability Criteria}
\label{sec2}

\begin{figure*}[t]
\centering
\includegraphics[width=1.5 \columnwidth]{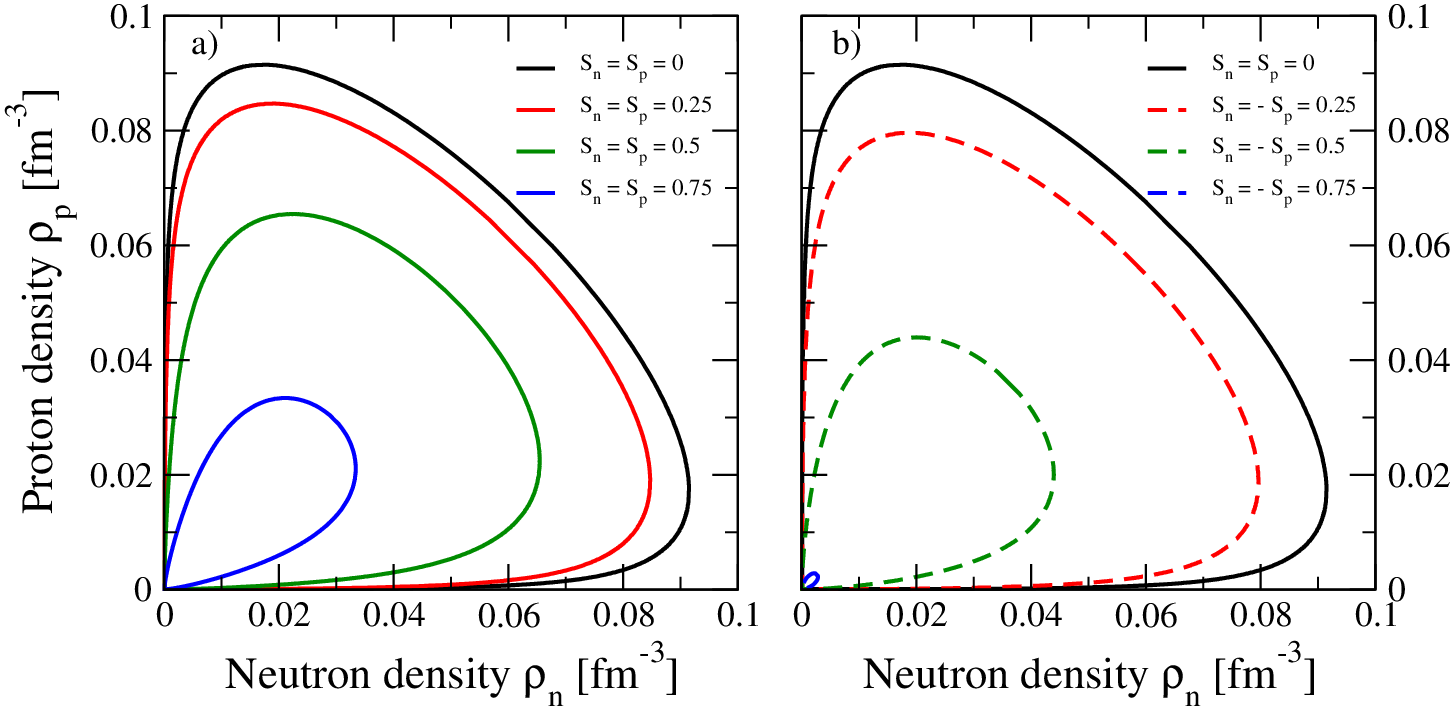}
        \caption{(Color online) Spinodal instability region for different combinations of the neutron and proton spin polarizations. Results for cases in which neutron and proton spins are oriented in a parallel and antiparallel way are shown in panels a and b, respectively.}
\label{fig:spinodal}
\end{figure*}

The stability of a system against a phase separation is guaranteed if the energy of a single-phase is lower than the energy of any multiple phase configurations. This condition is fulfilled if the free-energy density (energy density at zero temperature) is a convex function of the partial densities of the components of the system, that is, if the curvature matrix is positive definite.  In the case of spin polarized asymmetric nuclear matter, which is a 4 component system, the curvature matrix has a  $4 \times 4$ structure
\begin{equation}
C_{ij} = \left ( \frac {\partial ^2 \varepsilon}{\partial \rho_i \partial \rho_j} \right )\ , \,\,\,\, i,j= n_\uparrow, n_\downarrow, p_\uparrow, p_\downarrow \ .
\label{eq:2a}
\end{equation}
However, as we said in the introduction, in this work  we analyze the spinodal instabilities of polarized nuclear matter at fixed values of the neutron and proton spin polarizations 
$S_n$ and $S_p$. In this case, the thermodynamical stability against phase separation is guaranteed  by requiring  the convexity of energy density on its dependence of  the
total neutron ($\rho_n$) and proton ($\rho_p$) densities at given values of $S_n$ and $S_p$. The curvature matrix in this case is simply :
\begin{equation}
C=
\left(
\begin{array}{cc}
\frac {\partial ^2 \varepsilon}{\partial \rho_n^2} &  \frac {\partial ^2 \varepsilon}{\partial \rho_n \partial \rho_p}  \\
 \frac {\partial ^2 \varepsilon}{\partial \rho_p \partial \rho_n} & \frac {\partial ^2 \varepsilon}{\partial \rho_p^2}  
\end{array}
\right)_{S_n,S_p} \ .
\end{equation}

The condition of being positive defined requires that both the trace and the determinant of $C$ should be positive, {\it i.e.,} 
\begin{eqnarray}
\mbox {Tr} (C) = \lambda_+ +\lambda_- \ge 0 \nonumber \\
\mbox {Det} (C) = \lambda_+ \lambda_- \ge 0 \, , 
\label{eq:3}
\end{eqnarray}
where 
\begin{equation}
\lambda_{\pm} = \frac {1}{2} \left ( \mbox{Tr}(C) \pm \sqrt { (\mbox{Tr}(C ))^2 - 4 \mbox{Det} (C)} \right ) \, ,
\label{eq:4}
\end{equation}
are the two eigenvalues of the curvature matrix which have two associated eigenvectors $(\delta \rho_n^{\pm},\delta \rho_p^{\pm})$ with
\begin{equation}
\frac {\delta \rho_i^{\pm}}{\delta \rho_j^{\pm}} = \frac {\lambda_{\pm} - C_{jj}}{C_{ji}} , ~~~~i,j=p,n \,.
\end{equation}
Stability requires, that both eigenvalues should be positive.  It turns out that, for any fixed values of the neutron and proton spin polarizations $S_n$ and $S_p$,  $\lambda_+$ is always positive and only $\lambda_-$ can eventually become negative, signaling the beginning of the instability and the phase separation.  In addition, the magnitude of $\lambda_+$ exceeds always that of $\lambda_-$ ({\it i.e.,} $\lambda_+>|\lambda_-|$) and, therefore, the trace of the curvature matrix appears to be always positive. Consequently, the spinodal 
instability region for fixed values of the spin polarizations will be just determined by the values of the total neutron and proton  densities  which make the determinant of the curvature matrix negative as in the non-polarized case. 

\section{Results}
\label{sec3}

\begin{figure*}[t]
\centering
\includegraphics[width=1.5 \columnwidth]{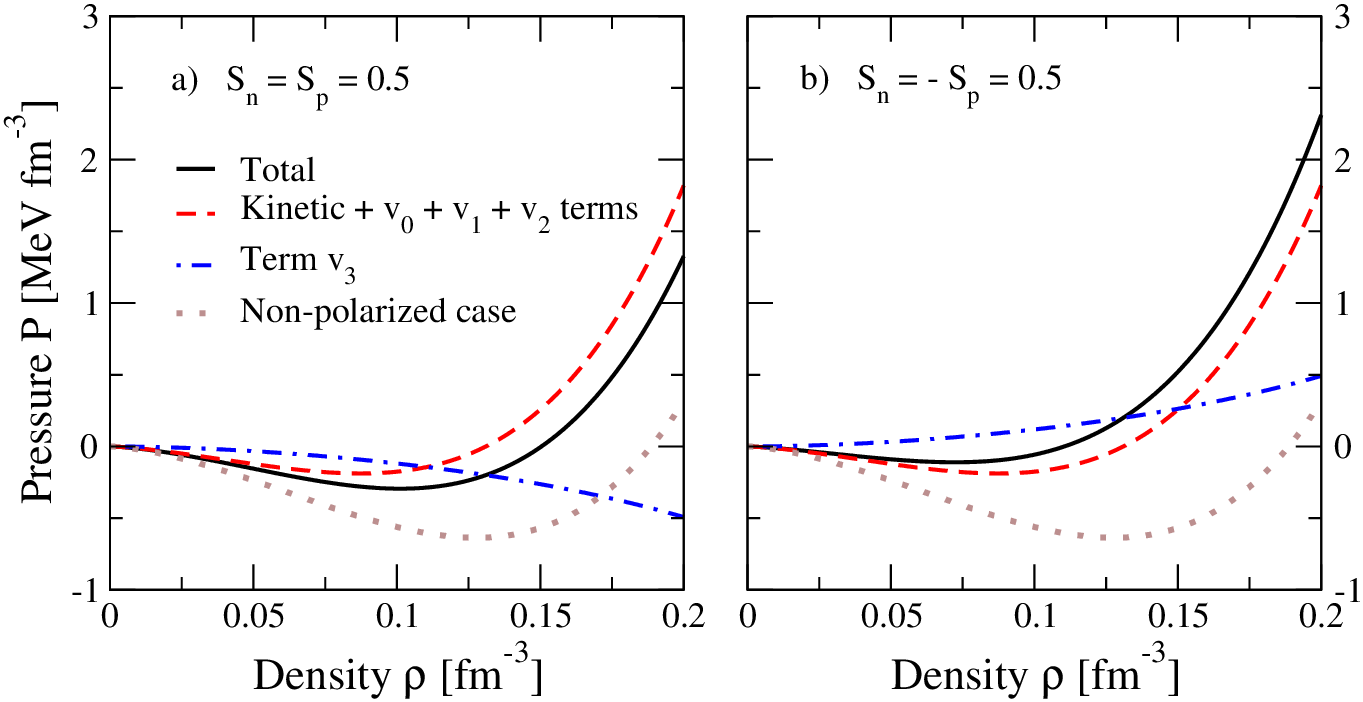}
        \caption{(Color online) Pressure of spin polarized  isospin symmetric nuclear matter. Results for $S_n=S_p=0.5$ are shown in panel a, whereas those for $S_n=-S_p=0.5$ are plotted in panel b. The separate contributions to the pressure of the kinetic energy density plus the terms multiplied by the coefficients  $v_0(\rho)$, $v_1(\rho)$ and $v_2(\rho)$ as well as that of the term multiplied by $v_3(\rho)$ of the parametrization of Eq.\ (\ref{eq:para}) are shown separately. The pressure for the non polarized case is also shown for comparison. }
\label{fig:press}
\end{figure*}

We start this section by showing in Fig. \ref{fig:spinodal} the spinodal instability region for different combinations of the neutron and proton spin polarizations.  We show results for the case in which the neutron and proton spins are aligned parallel to a given direction (panel a), and that in which they have an antiparallel orientation (panel b). To simplify the discussion, in all cases, we have considered that neutron and proton spin polarizations are the same in absolute value ($|S_n|=|S_p|$). As it can be seen, independently of the orientation of the spins,  the spinodal instability region shrinks when the system is polarized. We note that the instability region disappears completely when matter is totally polarized ($S_n=S_p=1$ and $S_n=-S_p=1$). We notice also that if the orientation of the neutron and protons spins is parallel the spinodal instability regions is always larger  than the one obtained when the spins are aligned in an antiparallel way. Note, for instance, that for $S_n=-S_p=0.75$ the region is extremely small, being almost completely suppressed, whereas for $S_n=S_p=0.75$  it is much larger, although its size is also clearly reduced with respect to the non-polarized case. 

We can understand the reduction of the spinodal instability region with the spin polarization by analyzing the behavior of the pressure
in the case of isospin symmetric nuclear matter, where the character of the spinodal instability is purely mechanical. Results for spin polarizations  $S_n=S_p=0.5$  and $S_n=-S_p=0.5$ are shown as an example in panels a and b of Fig.\ \ref{fig:press}, respectively. The pressure of the non-polarized case is also shown for comparison. As it can be seen, when the system is polarized its pressure  increases with respect to the non-polarized case, and the (mechanical) instability region (where the pressure derivative is negative) reduces. This is due, first, to the increase of the kinetic energy contribution to the pressure, which is always larger in the polarized system  (see Eq.\ (\ref{eq:kin})), and second, to the potential energy contribution, which in the polarized case, varies faster with density due to the spin polarization terms in the energy density functional of Eq.\ (\ref{eq:para}). Note, in particular, that  the derivative with respect to the density of the coefficient $v_2(\rho)$ is positive and larger, in absolute value, than that of $v_3(\rho)$ (see Fig.\ \ref{fig:coeff}). Therefore, the contribution from the spin polarization terms always increases the pressure when the system is polarized. Note also that the derivative of the coefficient $v_3(\rho)$ is negative. Hence, if the neutron and proton spins are oriented in a parallel (antiparallel) way the contribution to the pressure from the neutron-proton  spin polarization term will be negative (positive). As a result, the instability region will be larger (smaller) when  neutron and proton spins are parallel (antiparallel) oriented, as it can be seen in the figure. For completeness, we show also in the figure the  sum of the contribution to the pressure of the kinetic energy plus that of the terms $v_0(\rho)$, $v_1(\rho)$ and $v_2(\rho)$. As expected these contributions are the same independently of the orientation of the neutron and proton spins.  Similar conclusions can be drawn from the more cumbersome analysis of the curvature matrix.

\begin{figure*}[t]
\centering
\includegraphics[width=1.2 \columnwidth]{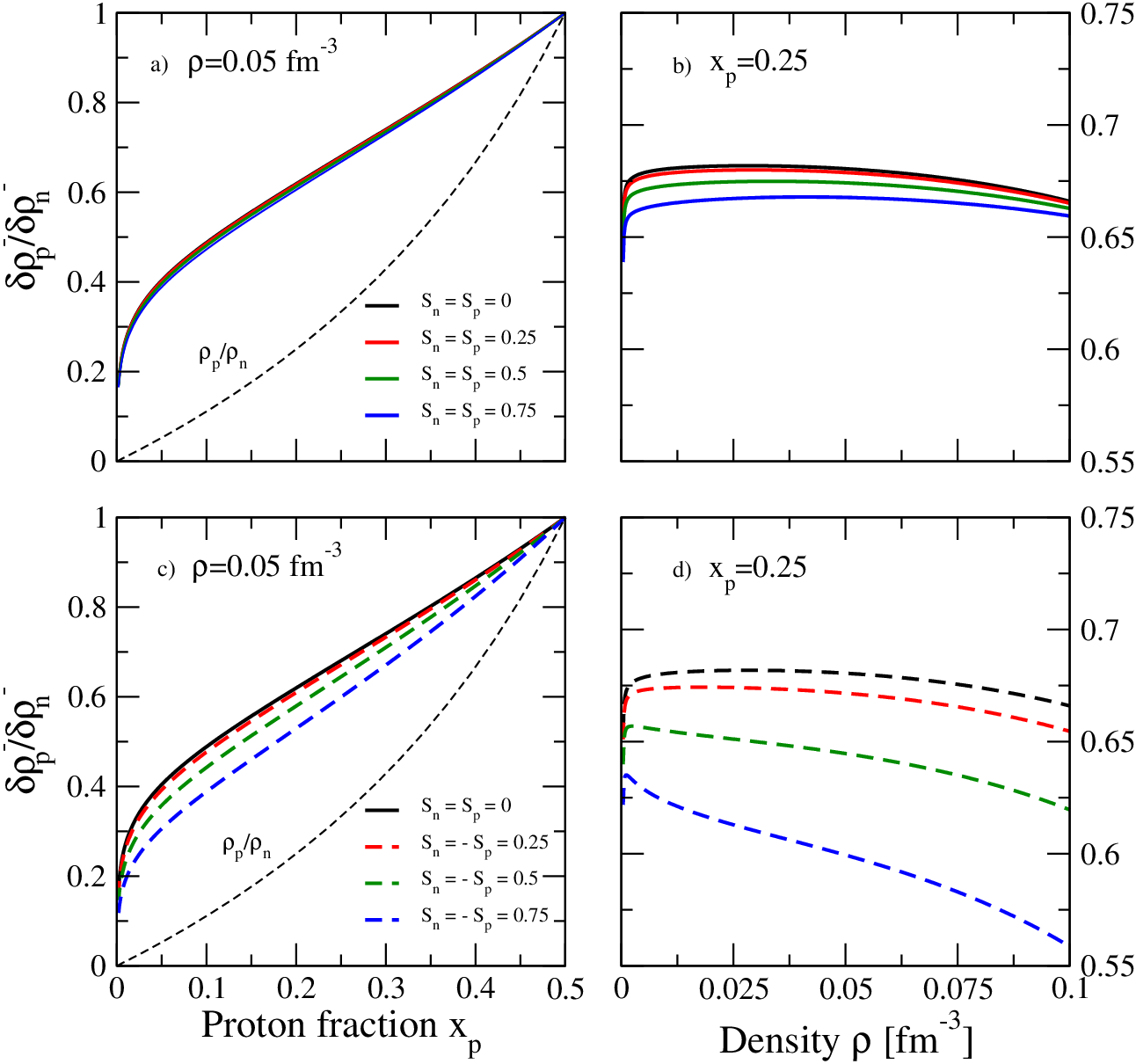}
        \caption{(Color on line) Left panels: ratio $\delta\rho^-_p/\delta\rho^-_n$ as a function of the proton fraction for  different combinations of neutron and protons spin polarizations at a fixed density $\rho=0.05$ fm$^{-3}$. Right panels: same ratio as a function of the total density for the same values of neutron and proton spin polarizations at fixed proton fraction $x_p=0.25$.  Results for the cases in which neutron and proton spins are oriented in a parallel (antiparallel) way  are shown in panels a (c) and b (d). The thin black dashed line shows the ratio between the protons and neutrons $\rho_p/\rho_n$.}
\label{fig:eigen}
\end{figure*}

As already pointed out in the introduction, asymmetric nuclear matter presents only one type of thermodynamical instability  \cite{baran01,margueron03,chomaz04} and not two independent ones (mechanical (or isoscalar) and chemical (or isovectorial)) as it has been usually argued. This instability appears, in fact, as a mixture of density and proton fraction fluctuations, and its direction is given by the ratio $\delta\rho^-_p/\delta\rho^-_n$ of the two components of the  eigenvector ($\delta\rho^-_n,\delta\rho^-_p)$ associated with the negative eigenvalue $\lambda_-$. This ratio tells us which is the predominantly character of the instability (isoscalar or isovector). Furthermore, it measures also the efficiency in restoring the isospin symmetry in the liquid phase. The larger its value, the greater the efficiency. In general, the nature of the instability will never be neither purely mechanical nor chemical, but it will appear as a mixture of both being predominantly of isoscalar type ({\it i.e.,} dominated by density fluctuations) if $\delta\rho^-_p/\delta\rho^-_n>0$ or of isovector type  ({\it i.e.,} dominated by proton fraction fluctuations ) if $\delta\rho^-_p/\delta\rho^-_n<0$. Only if  $\delta\rho^-_p/\delta\rho^-_n=\rho_p/\rho_n$ then the instability will preserve the ratio between protons and neutrons at which the system was prepared, and its nature will be purely mechanical, while if $\delta\rho^-_p=-\delta\rho^-_n$ then the total density of the system  will remain constant and, therefore, the instability will be purely chemical. 

We show in Fig.\ \ref{fig:eigen} the ratio $\delta\rho^-_p/\delta\rho^-_n$ as a function of the proton fraction (panels a and c) for a fixed density $\rho=0.05$ fm$^{-3}$, and as function of the density (panels b and d) at a fixed proton fraction $x_p=0.25$. Results for the cases in which neutron and proton spins are oriented in a parallel or antiparallel way are shown in the upper and lower panels, respectively, for the same values of $S_n$ and $S_p$ of Fig.\ \ref{fig:spinodal}. We note that in all cases $\delta\rho^-_p/\delta\rho^-_n$ is positive, indicating that, independently of the spin polarization, the instability is always dominated by total density fluctuations. 
Notice, however, that when the system is polarized the ratio $\delta\rho^-_p/\delta\rho^-_n$ decreases. This decrease is quite small when the neutron and proton spins are parallel and much larger if their orientation is antiparallel. Nevertheless,  the reduction of $\delta\rho^-_p/\delta\rho^-_n$  is not enough to modify the dominant isoscalar nature of the instability, which would be only signaled  by a change in the sign of the ratio. The decrease of $\delta\rho^-_p/\delta\rho^-_n$  indicates also that isospin symmetry restoration is less efficient when nuclear matter is polarized. We notice also that for symmetric matter ($x_p=0.5$)  $\delta\rho^-_p/\delta\rho^-_n=1$, indicating in this case that the instability occurs, as expected, in the pure isoscalar direction and that matter behaves as a one component system. Finally, we observe that $\delta\rho^-_p/\delta\rho^-_n$  is always larger than the ratio between protons and neutrons $\rho_p/\rho_n$ (see panels a and c). This is an indication that the instability drives the dense phase (liquid) of the system towards a more symmetric region in the $\rho_n-\rho_p$ plane. As a consequence, due to the conservation of the total number of particles, the light phase (gas) is enforced to become more neutron rich leading to the so-called isospin distillation or fragmentation effect \cite{xu00}. 



\section{Summary and Conclusions}
\label{conclusions}

In this work we have analyzed spinodal instabilities of  spin polarized asymmetric nuclear matter at zero temperature within the microscopic BHF approach using the Argonne V18 nucleon-nucleon potential plus a three-nucleon force of Urbana type. We have considered several configurations of the neutron and proton spins ranging from the non-polarized case to the totally polarized one.
Since BHF calculations are quite expensive in terms of computational time, to determine the spinodal instability region in a fast way we have employed an analytical parametrization of the energy density of spin polarized isospin asymmetric nuclear matter that reproduces with a good accuracy the microscopic BHF results. Our results have shown that independently of the orientation of neutron and proton spins, the spinodal instability region shrinks when the system is polarized, being its size smaller when neutron and proton spins are in an antiparallel way than when they are parallely oriented. Analyzing the pressure of spin polarized isospin symmetric nuclear matter we have found  that the reduction of the instability region 
in the polarized case with respect to the non-polarized one is due to: (i) the increase of the kinetic energy contribution to the pressure  which is always larger in the polarized system, and (ii) to the faster variation with density of the contributions to the pressure from the neutron-neutron, proton-proton and neutron-proton spin polarization terms.  We have found that it is in fact the neutron-proton spin polarization term the one that gives rise to the largest reduction of instability region if the neutron and proton spins are antiparallel. Finally, by analyzing the density and proton fraction dependence of the ratio $\delta\rho^-_p/\delta\rho^-_n$ we have found that, independently of the spin polarization, the spinodal instability is always dominated by total density fluctuations and that   $\delta\rho^-_p/\delta\rho^-_n$ decreases when the system is polarized, although this reduction is not enough to change the dominant isoscalar nature of the instability. We have also found that the restoration of the isospin symmetry in the liquid phase becomes less efficient with the polarization of the system. 

\vspace*{-0.3cm}
\begin{acknowledgments}
Isaac Vida\~na wants to express the deep gratitude he feels towards his mentor, colleague and, above all, friend Artur Polls who unexpectedly passed away  few weeks after this work was completed. 

\end{acknowledgments}


\end{document}